# A Local Fluctuation Theorem

*Gary Ayton , Denis J. Evans and D. J. Searles* ◊

Research School Of Chemistry

Australian National University

Canberra, ACT 0200 Australia

◊Department of Chemistry

University Of Queensland

Brisbane, Qld 4072 Australia

Monday, 25/1/1999

**Abstract**

The Fluctuation Theorem (FT) gives an analytic expression for the probability, in a nonequilibrium system of finite size observed for a finite time, that the dissipative flux will flow in the reverse direction to that required by the Second Law of Thermodynamics. In the present letter a Local version of the Fluctuation Theorem (LFT), is derived. We find that in the case of planar Poiseuille flow of a Newtonian fluid between thermostatted walls, non-equilibrium molecular dynamics simulation results support LFT.

PACS numbers: 05.20.-y, 05.70.Ln, 47.10.+g, 47.40.-n

In 1993 Evans, Cohen and Morriss [1], gave a formula for the logarithm of the probability ratio that in a nonequilibrium steady state, the time averaged entropy production per unit volume, takes on a value $\bar{\sigma}(t)$, to minus that value $-\bar{\sigma}(t)$:

$$\lim_{t\to\infty} \frac{k_B}{(t\bar{\sigma}(t)V)} \ln\left(\frac{p(+\bar{\sigma}(t))}{p(-\bar{\sigma}(t))}\right) = 1. \tag{1}$$

This formula has come to be known as the Fluctuation Theorem, FT. Surprisingly perhaps, it is valid far from equilibrium in the nonlinear response regime [1]. In 1994, Evans and Searles [2-4], gave a derivation, similar to that given here, which considered transient, rather than steady state, nonequilibrium averages and employed the Liouville measure. In 1995 Gallavotti and Cohen [5], gave a proof of (1) for a nonequilibrium steady state, based on a Chaotic Hypothesis and employing the SRB measure. Other generalizations of FT have been recently been developed [6,7].

FT gives an analytic expression for the probability that, for a finite system and for a finite time, the dissipative flux flows in the reverse direction to that required by the Second Law of Thermodynamics. It has been confirmed numerically [1-3, 7-10].

The generality of FT prompted Gallavotti to suggest in 1998 [11] that a local version of the theorem (LFT) might be valid, and recent studies have considered LFT [9, 12]. For macroscopic systems, the probability of observing Second Law violations is, as predicted by FT, unobservably small. If one wants to test FT using laboratory experiments (*e.g.* light scattering), a local version which applies to small subregions of a large macroscopic system would be extremely useful. In the present Letter we derive a new LFT on the basis of arguments from linear irreversible thermodynamics. We describe computer simulations which support the validity of LFT.

For an N-particle system in 3 Cartesian dimensions, with coordinates and momenta, $\{\mathbf{q}_1,\mathbf{q}_2,..\mathbf{q}_N,\mathbf{p}_1,..\mathbf{p}_N\} \equiv (\mathbf{q},\mathbf{p}) \equiv \Gamma$. The energy of the system is $H \equiv \sum_{i=1}^{N} p_i^2/2m_i + \Phi(\mathbf{q})$ where $\Phi(\mathbf{q})$ is the interparticle potential energy which is a function of the coordinates of all of the particles, $\mathbf{q}$. In the presence of an external field $\mathbf{i}F_e$, where $\mathbf{i}$ is a unit vector in the x-direction, the thermostatted equations of motion are taken to be,

$$\dot{\mathbf{q}}_i = \mathbf{p}_i / m_i, \quad \dot{\mathbf{p}}_i = \mathbf{F}_i(\mathbf{q}) + ic_i F_e - \alpha(\Gamma) S_i \mathbf{p}_i, \qquad (2)$$

where, $\mathbf{F}_i(\mathbf{q}) = -\partial \Phi(\mathbf{q}) / \partial \mathbf{q}_i$, $c_i = 0,1$, $S_i = 0,1$ so that $\sum_{i=1}^{N} S_i = N_w$ is the number of thermostatting particles, $\sum_{i=1}^{N} c_i = N_f$ the number of particles subject to the influence of $F_e$. $\alpha$ is the thermostat multiplier derived from Gauss' Principle of Least Constraint in order to fix the total energy [13]. The thermostat multiplier is $\alpha = -J(\Gamma) V F_e / 2K_W$, where $K_W = \sum_{i=1}^{N} S_i p_i^2 / 2m_i \equiv 3N_w k_B T_W / 2 \equiv 3N_w / 2\beta_W$, and the dissipative flux, J, is defined,

$$-J(\Gamma)V = \int d\mathbf{r} \sum_i [\frac{c_i p_{xi}}{m_i} \delta(\mathbf{r}_i - \mathbf{r})] \equiv \int d\mathbf{r}\, n(\mathbf{r}) u_x(\mathbf{r}). \qquad (3)$$

From previous work [3] we know that the Kawasaki-Lagrange form of N-particle distribution function, $f(\Gamma,t)$, is, $f(\Gamma(t),t) = \exp[-\int_0^t ds\, \beta_W J(s) V F_e] f(\Gamma(0),0)$ and from the Liouville equation [13] the change in the fine grained Gibbs entropy as a function of time is[1],

$$S(t) = -k_B \int d\Gamma\, f(\Gamma,t) \ln[f(\Gamma,t)] = S(0) - 3N_w k_B t \overline{\alpha}(t). \qquad (4)$$

We use the notation for a trajectory segment, $\Gamma(s); 0 < s < t$: $\overline{A}(t) \equiv \frac{1}{t}\int_0^t ds\, A(\Gamma(s))$.

Without loss of generality we assume that the initial t=0 ensemble is microcanonical. Since the dynamics is time reversible we know [2] that the logarithm of the ratio of probabilities,

$$\ln\left(\frac{p(-\overline{\beta_W J}(t))}{p(\overline{\beta_W J}(t))}\right) = \overline{\beta_W J}(t) V F_e t = -3N_w \overline{\alpha}(t) t. \qquad (5)$$

where $\overline{\beta_W J}(t) \equiv (1/t)\int_0^t ds\, 3N_w J(\Gamma(s))/2K_W(s)$. Unlike (1), no $\lim_{t \to \infty}$ is required for this version of FT. It is clear that the ratio, $p_-/p_+$, of observing an anti-trajectory rather than a trajectory is given by the *integrated* form of FT (IFT),

$$\frac{p_-(t)}{p_+(t)} = \left\langle \exp[-3N_w \overline{\alpha}(t) t] \right\rangle_+, \qquad (6)$$

---

[1] Note (4) is entirely consistent with thermodynamics: $dS/dt = (1/T_W)dQ/dt = -JVF_e/T_W = 3N_w k_B \alpha$.

where $\langle ... \rangle_+$ denotes an average over all transient trajectory segments for which $\bar{\alpha}(t)$ takes on a positive sign. These formulae are extremely general and are even valid in the far from equilibrium nonlinear regime [1-4]. In the long time limit, averages over *transient* segments which originate from the initial equilibrium microcanonical ensemble approach those taken over nonequilibrium *steady state* segments. Therefore (5, 6) are also true asymptotically, $\lim_{t \to \infty}$, in the nonequilibrium steady state.

FT and IFT are also valid for mixtures where different particles, i,j have different interparticle interactions. For our present purposes it is convenient to consider a mixture that is segregated into fluid and solid particles and where the solid particles form planar walls with a normal which is orthogonal to the applied field, $\mathbf{i}F_e$. We also assume that $c_i=1$, $S_i=0$ for all fluid particles, $c_i=0$ for all wall particles and $S_i=1$ for *some* wall particles far away from the fluid-solid interface - see Fig. 1. The system under study is gravity-driven planar Poiseuille flow of a fluid between planar thermostatted walls.

We assume our inhomogeneous system satisfies local thermodynamic equilibrium. For $\lim_{t \to \infty}$ there is a balance between $\bar{\sigma}(t)$, and the time averaged **total** entropy flux, $\bar{\mathbf{J}}_{ST}$

$$\lim_{t \to \infty} \int_V d\mathbf{r} \, \bar{\sigma}(\mathbf{r},t) = \lim_{t \to \infty} \int_S d\mathbf{S} \bullet \bar{\mathbf{J}}_{ST}(\mathbf{r},t), \tag{7}$$

where the volume V has an enclosing surface S with outward normal $d\mathbf{S}$ [13]. Dividing the entropy flux into the usual convective and diffusive components[2], $\mathbf{J}_{ST} = \rho \mathbf{u} s + \mathbf{J}_S$ $= \rho \mathbf{u} s + \mathbf{J}_Q / T$, where T is the local temperature, s the local entropy density and $\mathbf{J}_Q$ is the heat flux vector, we see that the total irreversible entropy production in a sub-region $[0, y_1]$ (y=0 is the centre of the flow channel) is

$$\dot{S}(0,y_1) = A \int_0^{y_1} dy \, \frac{\partial J_{Qy}(y)/T(y)}{\partial y} = A \frac{J_{Qy}(y_1)}{T(y_1)}. \tag{8}$$

A is the cross sectional area across which the heat flows. IFT for the fluid system $[-y_W, +y_W]$ can be written using (6) as

$$\frac{p_-(t)}{p_+(t)} = \left\langle \exp[-tA(\overline{\beta J}_{Qy}(y_w,t) - \overline{\beta J}_{Qy}(-y_w,t))]\right\rangle_+ = \left\langle \exp[-3N_w \,\overline{\alpha}(t)t]\right\rangle_+ \equiv \phi(t) \quad (9)$$

---

[2] Note that for our steady state geometry there are no contributions from the chemical potential and associated mass fluxes.

We note that the first equality in (9) does not refer at all to the thermostatting mechanism in the walls. Instead it only refers to the local heat flux and local temperature in a fluid which obeys purely Newtonian mechanics (*i.e.* no thermostats). Since the thermostatting particles may be *arbirarily far* from the fluid and since there is no way that the fluid particles can "know" exactly how the system is thermostatted at these distant walls, suggests a *local* FT (LFT),

$$\lim_{t\to\infty} \frac{p[-\overline{\beta J}_{Qy}(\pm y,t)]}{p[\overline{\beta J}_{Qy}(\pm y,t)]} = \lim_{t\to\infty} \exp[-At(\overline{\beta J}_{Qy}(y,t) - \overline{\beta J}_{Qy}(-y,t))], \quad (10)$$

and a *local* IFT (LIFT)[3]

$$\lim_{t\to\infty} \frac{p_-(\pm y,t)}{p_+(\pm y,t)} = \lim_{t\to\infty}\left\langle \exp[-At(\overline{\beta J}_{Qy}(y,t) - \overline{\beta J}_{Qy}(-y,t))]\right\rangle_+ \equiv \phi_L(y,t). \quad (11)$$

In deriving these equations we use the fact that $\overline{AB} = \overline{A}.\overline{B} + O(1/N)$ and that N is large. For example in (11) errors of $O(1/(n\xi \int_S dS))$ where $\xi$ is the correlation length in the fluid, are ignored. For a typical light scattering experiment these errors are insignificant, $O(1/10^6)$.

By integrating the time averaged conservation equations of hydrodynamics the heat flux can be calculated as[4],

$$\lim_{t\to\infty} \overline{J}_{Qy}(y,t) = \lim_{t\to\infty} F_e \int_0^y \overline{u}(y',t)\overline{n}(y',t)dy' - \lim_{t\to\infty} F_e \overline{u}(y,t)\int_0^y \overline{n}(y',t)dy'. \quad (12)$$

If the y coordinate extends to $y_w$ the second term vanishes and time averaging is not required when the entire system is ergostatted.

---

[3] The general form of LFT is, $\lim_{t\to\infty} \dfrac{p[-\int_S d\mathbf{S} \bullet \overline{\beta \mathbf{J}}_Q(\mathbf{r},t)]}{p[\int_S d\mathbf{S} \bullet \overline{\beta \mathbf{J}}_Q(\mathbf{r},t)]} = \lim_{t\to\infty} \exp[-\int_S d\mathbf{S} \bullet \overline{\beta \mathbf{J}}_Q(\mathbf{r},t)t]$.

[4] Care is required in strongly inhomogeneous systems - see [14].

An NEMD algorithm was developed where fluid particles interacted via Newtonian dynamics, and the walls not only removed viscous heat, but created drag. The potential was chosen $\Phi(\mathbf{q}) = \sum_{i<j} \phi(q_{ij}) + (1-c_i)(1-c_j)\phi_{spr}(q_{ij})$ with, $\phi(q_{ij}) = 4(q_{ij}^{-12} - q_{ij}^{-6}) + 1, q_{ij} \leq 2^{1/6}$; $\phi(q_{ij}) = 0, q_{ij} > 2^{1/6}$, and polymer cross-links formed with $\phi_{spr}(q_{ij}) = -kq_{ij}^2$. k is a spring constant typical for a cross-linked polymer. An initial polymer was constructed from an equilibrium fluid ($\phi_{spr} = 0$) at a density $n = N/V = 0.8442$ and temperature $T = 0.722$. All pairs of polymer particles with $q_{ij} \leq 1.25$ were "linked", forming a cross-linked polymer. The polymer particles near the fluid-polymer interface evolved under Newtonian dynamics. However, particles in the ergostatting core of the polymer had their equations of motion supplemented by Gaussian forces which constrained to total momentum of these particles to be zero and the *total* energy of the *system* to be constant. Details are described in [15].

In the simulations N=798, $N_F$=320, $N_W$=240; the unit cell was, $L_x = L_z = 6.718$, $L_y = 20.154$. An equilibration run of $5 \times 10^5$ time steps was performed before production runs of $5\text{-}10 \times 10^6$ time steps ($\delta t = 0.001$). The polymer region had a width $L_{pol} = 12$, and under 3 dimensional periodic boundary conditions, each polymer surface formed an interface for the single fluid channel. The ergostatting core had a width $L_{erg} = 6$ - see Fig. 1.

We begin the discussion of the results with Fig. 1 where we show the average heat flux $J_{Qy}(y)$ across the channel for $F_e = 0.01$. Note that in the Newtonian polymer region, where $\partial u_x / \partial y = 0$ and $c_i = 0$, $J_{Qy}(y)$ plateaus. Also, as shown in the inset, the nearly constant temperature profile across the channel indicates that we are in the weak field linear response regime.

To test our expression (11), four different regions of increasing size, each symmetric about y=0, were used to calculate $J_{Qy}(y)A/T$: region 1 was $y \in [-.8,+.8]$ with its data shown as ◊ in Fig. 2, 2 was $y \in [-1.6,+1.6]$ shown ο, 4 was $y \in [-3.2,+3.2]$ shown ❏ [5] and 5 was $y \in [-4.0,+4.0]$ shown Δ. These regions are shown in Fig. 1.

We can examine the validity of LIFT (11) for the $F_e = 0.01$ system as shown in Fig. 2. Each pair of curves (designated by open and closed symbols) corresponds to $p_- / p_+$ and $\phi_L(t)$ respectively, evaluated in the different regions, where the most rapidly decaying pair corresponds to the whole system (global IFT) then region 5, region 4, 2, and 1, the slowest decaying region. Those regions nearest the interface exhibit LIFT similar to the global IFT. LIFT for the central fluid regions (2 and 1) decay more slowly and approach zero entropy production at the center of the channel. In each case the IFT (11), is confirmed at sufficiently long averaging times.

---

[5] To reduce clutter region 3 is not displayed.

In conclustion we have derived a LFT, which is valid for natural systems in local thermodynamic equilibrium. The theorem is valid for natural systems because although a mathematical device (Gaussian thermostat) is employed in order to create a nonequilibrium steady state, the region over which the theorem is applied is *not* subject to thermostatting. The thermostatting only occurs at remote thermal boundaries.

In the case of planar Poiseuille flow, we used hydrodynamic integration [14], (12) to calculate the local spontaneous entropy production and have verified numerically that sub-regions obey LIFT. We have thus shown that systems evolving under Newtonian mechanics (*i.e.* no thermostats), but in contact with a thermal reservoir obey LIFT, despite the fact that there is no *explicit* phase space compression in the region of the system under consideration.

The thermostatting region can be made arbitrarily remote from the fluid. This means that the mathematical details of the thermostatting mechanism cannot possibly affect the statistics of the entropy fluctuations in the fluid. The thermostat we use is a convenient but ultimately irrelevant mathematical device.

In their original paper [1], Evans et al. pointed out in a footnote that FT can be derived from the linear response Green-Kubo (GK) relations together with the Central Limit Theorem applied to $\{\overline{\sigma}(t)\}$. Since GK relations are independent of the thermostatting mechanism [13] it should come as no surprise that FT in the linear regime is robust with respect to the thermostatting mechanism.

Further work needs to be carried out to discover whether LFT and LIFT are valid in the nonlinear response regime. This is difficult to test because in the nonlinear regime, the fluxes must be sufficiently strong to observe a nonlinear response but sufficiently weak that reverse fluxes can be observed.

A version of (6) is known to be valid even for very small systems (N~8) [3]. It is not known whether LFT and LIFT (10, 11) are valid for similarly small systems.

**Acknowledgements**

We would like to thank the Australian Research Council for the support of this project. Detailed discussions and comments from Professor E.G.D. Cohen, are also gratefully acknowledged.

**Figures**

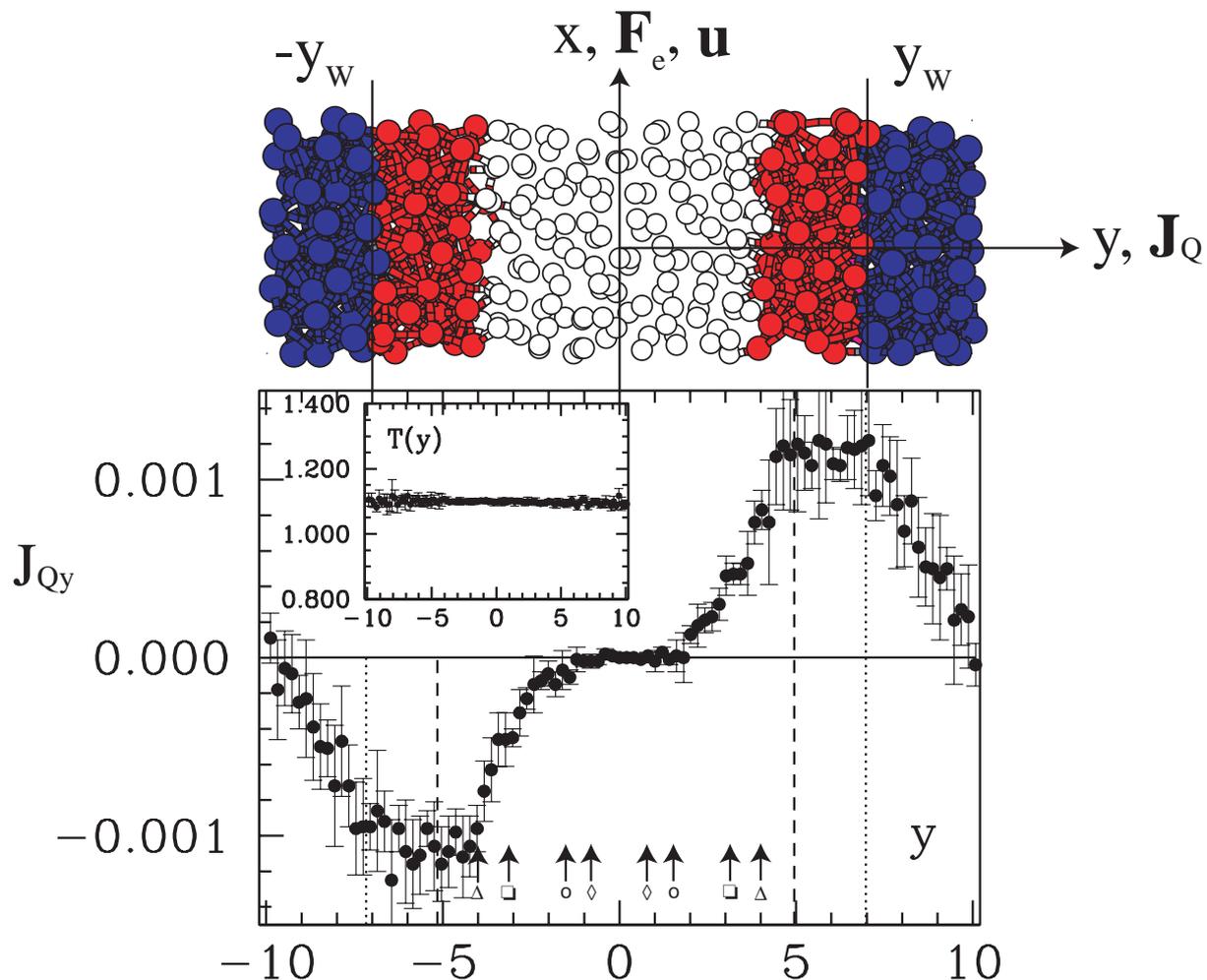

Fig.1 This two part figure shows at top, a randomly chosen, steady state snapshot of the atomic configuration of the system. The shaded particles comprise the cross-linked polymer walls, the darkest of these particles are the thermostatting polymer particles. The Newtonian fluid atoms are shown as white. On the same y-scale, the bottom part of this figure shows the average heat flux $J_{Qy}(y)$ for $F_e = 0.01$.

12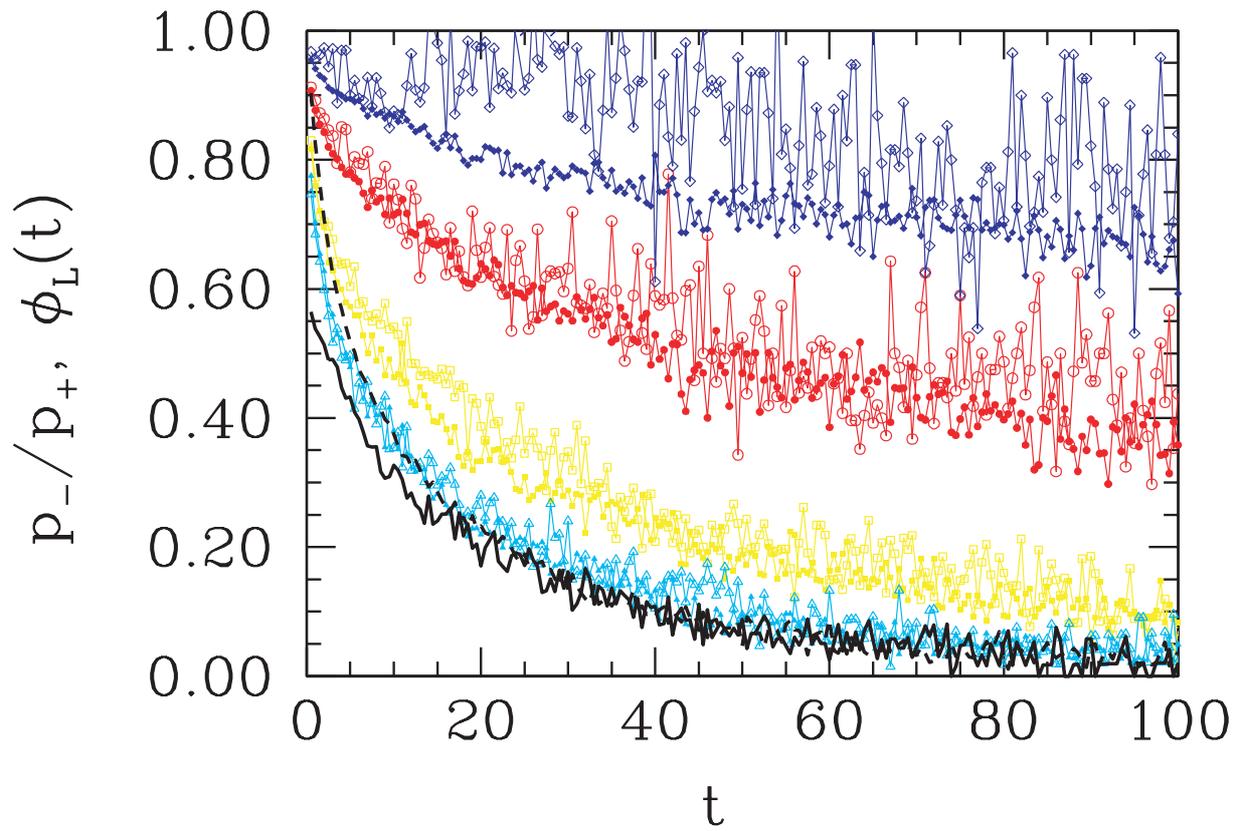

Fig.2. Compares $p_-/p_+$ and $\phi_L(t)$ for the 4 different regions. The solid curves at the bottom show the global data for the system (9).